# Addicts without Substance? Social Media Addiction when Facebook Shuts Down

*Short Paper*

**Darshana Sedera**
Southern Cross University
Gold Coast, Australia
darshana.sedera@gmail.com

**Sachithra Lokuge**
RMIT University
Melbourne, Australia
ksplokuge@gmail.com

## Abstract

*In March 2018, a series of anti-social and racial riots in Sri Lanka led to a government-controlled ban of all social media use in the country for 14 days. This nation-wide ban included the use of all social media such as Facebook, Twitter and communication apps like WhatsApp, Viber and WeChat. Until the day of the sanctions, a population of 23 million in Sri Lanka had never experienced government sanctions, restrictions or interventions on social media use. This sudden ban provided a unique window of opportunity to investigate social media non-use and use and how it might lead to psychological distress. Using a longitudinal study design of two surveys, analyzing data of 476 and 205 respectively, this study makes insightful preliminary observations of social media non-use and use continuum.*

**Keywords:** Social media, addiction, non-use, use, longitudinal survey

## Introduction

The role of social media on individuals has been studied in management (Wang et al. 2013), psychology (Pempek et al. 2009), information systems (Palekar et al. 2015; Sedera et al. 2017a), political science (Tufekci and Wilson 2012) and organizational environment (Arvidsson and Holmström 2013). Such literature demonstrates the imperative role that it plays in one's life in building social capital, managing self-presentation and enhanced communication with friends (Ellison et al. 2006). It is evident that the social media's substantial proliferation, ubiquity, richness, immediate gratification and anonymity, have made it a part of our daily routines (Alarifi et al. 2015; Leonardi 2014). With an estimated subscription base of 4 billion users for Facebook and Twitter in 2019, and with more than half of them logging in daily (Statistca 2017), social scientists believe that social media is fast becoming an addiction (Van Den Eijnden et al. 2016). Recently, researchers have associated social media use with several psychiatric disorders, including depression, anxiety and low self-esteem (Pantic 2014; Sedera et al. 2017b). Such studies argue that compulsive social media use or social media addiction is a growing mental health problem (Van Den Eijnden et al. 2016), drawing similarities with online gaming addiction (Van Rooij et al. 2011) and internet gambling addiction (King et al. 2014).

So...*what happens to social media addiction when the user is barred from using social media?* Building on the unique events that occurred in Sri Lanka in March 2018, this study makes several observations on how mandated social media *non-use* affects those regular social media users. Between 7th to 18th March 2018, *all* social media (e.g., Facebook, Twitter, Instagram) and *all* communication applications (e.g., WhatsApp, Viber) were blocked by the Government of Sri Lanka. The government mandated these unprecedented steps to curtail the spread of rumors surrounding racial tensions spreading from one local government area to the other 340 local government areas of the country. Prior to this incident, there had not been any restrictions to use social media or communication applications in Sri Lanka. Therefore, this background provided an opportunity to make longitudinal assessments of mandated non-use of social media and its impact on social media users.

As such, this research-in-progress paper focuses on social media use as an addiction, making comparisons between the mandate non-use period against the regular use period using a longitudinal survey. The paper proceeds in the following manner. First, it provides the background of the study. Next, it provides an overview of the longitudinal study design, the theoretical scaffolding and the research model. Finally, the initial findings are reported, followed by conclusions and limitations.

## Review of Related Work

### *Social Media Non-use*

Prior research highlights that individuals discontinue social media use as a statement of political or social identity, or as an example of an authentic human behavior (Portwood-Stacer 2013). In such studies, the user makes a voluntary and somewhat inspired step to discontinue social media use. Such aspirational non-use parallels with some social media users who had aspired to deactivate their social media accounts (Allcott et al. 2020). Related to this, researchers have observed motivations for social media non-use (Baumer et al. 2013) and the underlying demographic or behavioral differences between the communities of the network's use and non-use (Hargittai and Walejko 2008). Then, there are studies that made observations of *voluntary* periodic absence of social media non-use. For example, forgoing social media use during Lent, Schoenebeck (2014) observed that only 64% of users who proclaim that they are giving up Twitter for Lent, successfully did so. Similarly, Baumer et al. (2013) identified that nearly half of the respondents who left Facebook for a short period, subsequently returned to the site. Brubaker et al. (2016) focusing on Grindr– claimed to be the world's largest social networking app for gay, bi, trans, and queer people – found that social media non-use is a reversible process due to its addictive nature. This study, embedded in the context described above, provides a unique perspective to further our understanding of social media non-use. It observes how the '*mandated social media non-use,*' where an individual is legally prohibited from using social media, influences one's normality of life. As Brubaker et al. (2016, p. 14), rightfully state "research agendas around non-use may benefit from studying returns [and] cyclical adoption and departure."

### *Non-user Typologies*

Over the past several decades, system non-use has received substantial attention in the disciplines of human computer interaction (Satchell and Dourish 2009), psychology of technology (Lazo et al. 1997) and the sociology of technology (Wyatt 2003). For identifying non-users, Wyatt (2003) proposed a two-by-two typology based on two dimensions: 'volitionality' (whether non-use is voluntary?) and temporality (was the individual previously a user?). The quadrants of the typology include (i) *the rejecters* – those who previously used a technology, but voluntarily gave it up, (ii) *the resisters* – those individuals who had never used it, (iii) *the excluded* are users, against their will, prevented from using a technology, while (iv) *the expelled* previously used it, but then were forced to stop[1]. Our context and thus the sample bear the characteristics of the 'expelled,' due to the ban of the government. In this study we *introduce a new category* that we term as *'rejoicer.'* The rejoicer enjoys the transitions from the state of 'expelled' to the regular use of social media once the social media bans are over.

### *Social Media Addiction*

In recent years, there are growing evidences that social media use is turning into an addiction, particularly for adolescents (Pantic 2014)[2]. Following the definition of Internet addiction, social media addiction has been described as a state where an individual has lost control of the social media use and excessively use social media to the point, where he/she experiences problematic outcomes that negatively affects his/her life (based on the foundational works of Young 1998; Young and De Abreu 2011). Researchers highlight that lack of self-esteem and life satisfaction have a moderating effect on social media addiction (Hawi and Samaha 2017). Further, the extent of social connectedness makes it harder for individuals to be without it (Savci and Aysan 2017; Sedera and Lokuge 2018). To assess social media addiction, Van Den Eijnden et al. (2016) proposed a scale, which is now recognized by the Diagnostic and Statistical Manual of Mental Disorders (DSM-5). Employing their scale, studies have established the relationship between social media addiction and psychological distress, and highlighted that its effects surpass both age and gender (Baumgartner et al. 2018). Furthermore, specifically related to the study background of riots and racial tensions, researchers have highlighted that social media has the propensity to alleviate dysphoric moods and may therefore be used as a coping mechanism or as a compensative behavior for real life problems (Kuss et al. 2018). This notion is similar to compensatory potential of media use, which claimed that people were more likely to engage in bouts of heavy TV watching when they were in dysphoric states (Kubey and Csikszentmihalyi 2014).

---

[1] It is noted that others have further refined classifications such as *the lagging resister* (Baumer et al., 2013) who has strongly considered technology non-use, but not yet actually done so, and (iv) *reverter*, a rejecter who later becomes a user again.

[2] The addiction literature has extensively reflected on the existence of *non-substance related* or *behavioral addictions*, such as Internet addiction, Internet pornography addiction and Internet gaming addiction.

## Theory, Research Model and Research Design

Considering the research context, where social media is banned for the first time, during a time of riots and racial tensions, this research employs the *compensated internet use theory* (Kardefelt-Winther 2014). "*The basic tenet of the theory of compensatory internet use is that the locus of the problem is a reaction by the individual to his negative life situation, facilitated by an internet application*" (Kardefelt-Winther 2014, p. 352). As such, this theory provides an ideal scaffolding for the study. Moreover, this theory has been specifically designed to gauge the level of addiction to Internet and social media, with a strong dependent variable of psychological distress (for a review of such works, refer (Sowislo and Orth 2013)). Therein, the theory argues that, due to lack of social stimulation in one's real life, the Internet user reacts with a motivation to go online to socialize, which is facilitated by social media where socializing is afforded. When considering the behavior of the individual during the social media lockdown, the user will demonstrate the characteristics of *the expelled*, while after the lockdown, the user will demonstrate the characteristic of the newly conceptualized category of non-use to use transition: *the rejoicer*.

The research model in Figure 1 is derived using the compensated internet use theory. First, it highlights the two phases of measurement ($t_1$ and $t_2$) and the conceptualization of the user types. Second, the model conceives a relationship between social media addiction and psychological distress during bans and after the bans are lifted. Considering the theoretical structures, during period $t_1$, social media *non-use* period, the individual is likely to depict withdrawal symptoms of social media addiction. Whereas, for the period $t_2$ of social media *use*, the user is purported to demonstrate characteristics of social media addiction. Third, following the foundations of social impact theory (Latané 1996), three moderating variables: frequency of social media use, number of friends in social media and the number of frequently used social media applications were used to assess the impact of social media addiction leading to psychological distress (Palekar et al. 2015). The variable social media addiction was measured using the 9 items of Van Den Eijnden et al. (2016) and the variable psychological distress was assessed by Lovibond and Lovibond (1995) measures of psychological distress of depression, anxiety, stress, and anger. In addition, three criterion items were employed as global measures as per the guidelines (Gable and Sedera 2009).

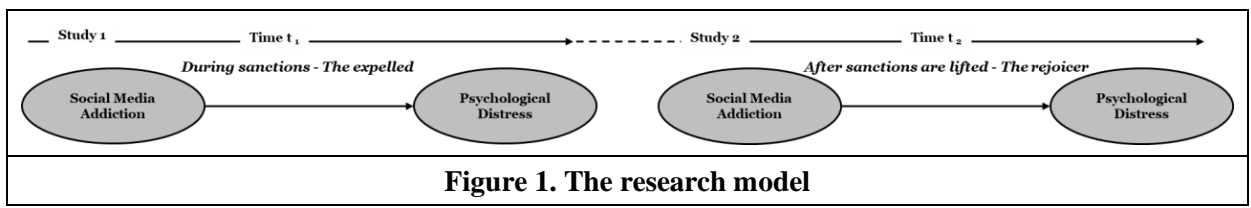

**Figure 1. The research model**

The questionnaire was pilot tested with 10 individuals who were not residing in Sri Lanka. This was intentional as to increase the face validity and generalizability, where a 'non-compensated' participant can relate to the questions and instructions of the survey instrument better. The pilot exercise particularly focused on improving the clarity and suitability of the measures. The consensus of the pilot panel with respect to all constructs and the measures suggested that the measurement instrument scales had adequate face validity. All the items were measured on a seven-point Likert scale with the end values of (1) 'Strongly Disagree' and (7) 'Strongly Agree,' and the middle value of (4) 'Neutral.' The study employed a convenience stratified sampling technique to gather data using a paper-based survey instrument. This allowed us to retain a balanced sample of age and gender, consistent of the intent and the design of the study. Moreover, no data was collected from the local government area of 'unrest,' which was marked by a police curfew. This allowed to minimize the potential effects of psychological distress arising from the events of political unrest – rather than from social media non-use. The social media landscape in Sri Lanka is consistent with most countries in the world, where Facebook dominates ~75% of the market, followed by Twitter and Instagram (StatCounter 2020). Such characteristics of the sample increases the generalizability of the findings.

The study design is a longitudinal one that includes data from the two phases: *during* and *after* the social media lockdown period. The first survey (Study-1) was conducted three days after the lockdown had commenced and the second survey (Study-2) was conducted three day after the lockdown was lifted. The study received a representative sample of 476 responses in total, of which 205[3] respondents had responded to both Study-1 and 2, making it possible to make matched longitudinal analysis. All items from all scales were subjected to an analysis of missing values in data, revealing less than 1.7% missing data for any included measures. When missing values were replaced using multiple imputation, no substantive changes in descriptive statistics were observed on any variable.

## The Preliminary Analysis

We employed partial least squares (PLS) structural equation modelling (SEM) method in the study. The analysis followed the guidelines of Benitez et al. (2018). The model validation conducted in 6 steps is reported below. To test the measurement and structural models, we employed ADANCO 2.0.1 software (Dijkstra and Henseler 2015) with the bootstrap resampling method (4999 resamples). The results of Study-2 are reported using square brackets.

The assessment of the measurement model commenced with a confirmatory composite analysis, which "checks the adequacy of the factor and composite models by comparing the empirical correlation matrix with the model-implied correlation matrix by examining the standardized root mean squared residual (SRMR), unweighted least squares (ULS) discrepancy ($d_{ULS}$), and geodesic discrepancy ($d_G$) for the saturated model" (Benitez-Amado, Henseler and Castillo, 2017, p 5). The SRMR of the model was 0.029 [Study-2 = 0.031], below the recommended threshold of less than 0.080 at the 0.05 alpha level (Dijkstra and Henseler 2015), with $d_{ULS}$ =0.154 and $d_G$ = 0.019 demonstrating that we can ensure with a probability of 5% that the measurement structure of the composite constructs are correct. As such, it is possible to proceed to evaluate the specific properties of the composite constructs.

All the constructs demonstrated satisfactory convergent and discriminant validity, with AVE for all variables measuring above 0.5 (Cenfetelli and Bassellier 2009). The AVE of each variable was greater than the variance shared between the construct and the other constructs in the model, indicating strong discriminant validity.

We tested for possible effects of the common method variance (CMV) using (i) Harman's single factor test (Harman 1976), with largest factor accounted for 27% [Study-2 = 31%] of the variance; and rotated factor loading matrix showing that the items for each construct loaded on a single factor (while items for different constructs loaded on different factors); and (ii) confirmed that none of the significance levels of correlations among independent and dependent variables changed when we partial out common method bias using an unrelated "marker variable" (Lindell and Whitney 2001). As such, we do not believe that common method bias is a serious issue.

Figure 2 depicts the path coefficient for Study-1 was 0.762 [Study-2 = 0.603] and the $R^2$ of the exogenous variable in Study-1 was 0.43 [Study-2 = 0.27]. Not only did the results of the nomological network testing (Figure 2) evidence the existence of a strong, positive and significant relationship between social media addiction and psychological distress as hypothesized, they further evidenced the aggressiveness of the psychological distress in the non-use period. To further assess the validity of the model, the research model was tested using the *common* respondents of both Study-1 and Study-2. Therein, the same structural model was tested using n = 205. The results indicated a path coefficient for Study-1 of 0.744 [Study-2 = 0.581] and the $R^2$ of the exogenous variable in Study-1 was 0.41 [Study-2 = 0.23]. The comparatively similar results in the analysis between the *common* sample and the two independent samples (Study-1 and 2), further encouraged the consistency and the generalizability of study results. To the best of our knowledge, this is the first study that investigated the psychological distress of social media addiction, when the social media use is withheld without consent. As such, the results established that social media addiction has a positive significant effect on psychological distress, explaining 43% of its variance in the non-use period (Study-1) and 37% during the use period (Study-2). Moreover, to the extent that the models are distinct and agree with the hypotheses, they support the existence of non-user category of 'expelled' in $t_1$ and the 'rejoicer' in $t_2$.

---

[3] By including their email address, the Study-1 included an option for the participants to take part in Study-2. As such, 414 (87%) of the sample in Study-1 provided their consent to take part in Study-2 (~50% response rate for Study-2).

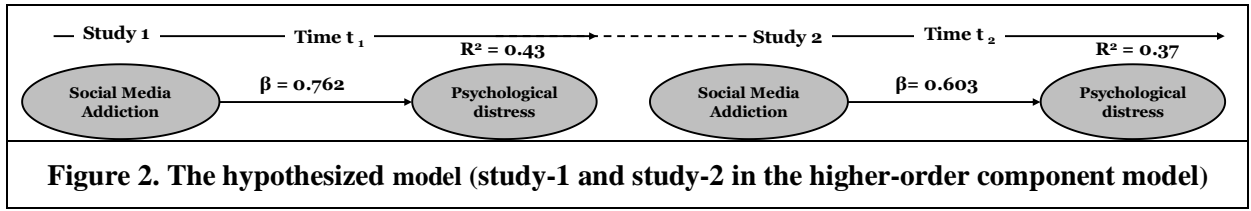

**Figure 2. The hypothesized model (study-1 and study-2 in the higher-order component model)**

A series of multi-group analyses (See Table 1) were conducted. In doing so, we commenced with the MICOM (measurement invariance of composite models) procedure, which assessed the invariance (Henseler et al. 2016). The results confirm the three types of invariance, which implied that measurement invariance holds and that a multi-group analysis is therefore possible (Hair et al. 2018).

| Table 1: Multi-Group Analyses | | | | | | | |
|---|---|---|---|---|---|---|---|
| Number of social media apps | | G 1 | G2 | Group 1 vs. Group 2 | | | |
| | | β | β | \| (β g1 - β g2) \| | t | Sig | p |
| Study-1 (G1 <3 = 278, G2 >3 = 150) | SM Addiction >> Psyc. Distress | 0.636 | 0.811 | 0.175 | 8.639 | Sig | 0.001 |
| Study-2 (G1 <3 = 111, G2 >3 = 74) | SM Addiction >> Psyc. Distress | 0.602 | 0.747 | 0.145 | 0.714 | NS | 0.426 |
| Number of friends in social media | | G 1 | G2 | Group 1 vs. Group 2 | | | |
| | | β | β | \| (β g1 - β g2) \| | t | Sig | p |
| Study-1 (G 1 = 234, G 2 = 252) | SM Addiction >> Psyc. Distress | 0.633 | 0.788 | 0.155 | 7.341 | Sig | 0.001 |
| Study-2 (G 1 = 105, G 2 =100) | SM Addiction >> Psyc. Distress | 0.571 | 0.801 | 0.23 | 10.016 | Sig | 0.001 |
| Frequency of regular use | | G 1 | G2 | Group 1 vs. Group 2 | | | |
| | | β | β | \| (β g1 - β g2) \| | t | Sig | p |
| Study-1 (G1 Low=162, G2 Heavy=314) | SM Addiction >> Psyc. Distress | 0.539 | 0.753 | 0.214 | 9.562 | Sig | 0.001 |
| Study-2 (G1 Low=70, G2 Heavy=135) | SM Addiction >> Psyc. Distress | 0.556 | 0.661 | 0.105 | 0.641 | NS | 0.528 |
| Age | | G 1 | G2 | Group 1 vs. Group 2 | | | |
| | | β | β | \| (β g1 - β g2) \| | t | Sig | p |
| Study-1 (Millen = 200, Gen-Y = 276) | SM Addiction >> Psyc. Distress | 0.637 | 0.643 | 0.006 | 0.502 | NS | 0.504 |
| Study-2 (Millenn = 92, Gen-Y = 113) | SM Addiction >> Psyc. Distress | 0.664 | 0.549 | 0.115 | 0.653 | NS | 0.554 |
| Gender | | G 1 | G2 | Group 1 vs. Group 2 | | | |
| | | β | β | \| (β g1 - β g2) \| | t | Sig | p |
| Study-1 (M= 248, F = 228) | SM Addiction >> Psyc. Distress | 0.72 | 0.805 | 0.085 | 3.199 | Sig | 0.001 |
| Study-2 (M = 107, F = 98) | SM Addiction >> Psyc. Distress | 0.622 | 0.771 | 0.149 | 1.794 | Sig | 0.001 |

**Table 1. Multi-Group Analyses**

Two groups were created based on the number of social media apps an individual regularly uses. In the non-use phase (Study-1), results indicated that there are significant differences between those who use less than two apps versus those who use more. Such differences are not significant with Study-2. We argue that the absence of social media endured during non-use phase, while substantial for all, the effect augments with the number of social media apps regularly used by an individual.

Using a globally accepted average of 300 friends in Facebook (Statistca 2017), two sub-samples, one with less-than 300 friends and the second with more than 300 friends were created. The analyses showed that both sub-samples resulted in significant results. However, interestingly, we noted that the path coefficient is the highest for Study-2 (see Table 1), when bans are lifted. This is *inconsistent* with our expectation for individuals during the period of social media non-use.

Two sub-samples were created to assess whether the frequency of use (i.e. low and heavy use of social media, employing two Likert items, with values over 4 considered heavy and below considered low). Results indicated that these two groups demonstrate significant differences for both Study-1 and Study-2. However, the high β values for the heavy users *during* the social media ban highlight that, heavy social media users feel higher psychological distress compared to the less frequent users. Results show that the (i) distress occurs through homeostatic imbalance occurred through the ban and (ii) that the distress is greater when imbalance is greater (loss of more when use was high against low use).

Two sub-samples were derived based on age – millennials (born post 1980) and Gen-Y (early-to-mid 1960s to 1980s). The high β values for both sub-samples were noted. However, the comparisons did not demonstrate significant differences, alluding to the possibility that social media addiction leading to psychological distress is prevalent amongst all age groups.

Exploring the gender differences of social media addiction leading to psychological distress, the two sub-samples of male and female did not demonstrate significant results for either period. However, it was observed that female cohort demonstrated higher β in both study periods, compared to their male counterparts. This is inconsistent with studies like Labad et al. (2008), who found that gender did not make a difference in psychological disorders. However, this is consistent with digital addiction studies (Wheaton et al. 2008), in which they reported that the symptoms of digital disorders are severe amongst women.

### *Longitudinal Assessment*

It is important to assess how the addiction or psychological distress revealed during the non-use carries over to the phase where the social media use is not constrained. As such, we conducted an analysis to assess the carry-over effect, where the same respondents answered to the same questions in different points in time. Herein, we only employed the panel data of the 205 respondents, who had responded for both Study-1 and Study-2.

| Table 2. Longitudinal Assessment | | | | | | |
|---|---|---|---|---|---|---|
| **Type** | **Time** | **Effect** | **Path Coefficient** | **t-value** | **Sig** | **p-value** |
| Direct Effect | t1 | SMA (t1) > PD (t1) | 0.762 | 14.375 | Sig | 0.001 |
| | t2 | SMA (t2) > PD (t2) | 0.603 | 13.881 | Sig | 0.001 |
| Carry-over effect | t1/t2 | SMA (t1) > SMA (t2) | 0.801 | 16.11 | Sig | 0.001 |
| | t1/t2 | PD (t1) > PD (t2) | 0.433 | 7.594 | Sig | 0.001 |
| Multi-Group Analysis (Study-1 n = 205, Study-2 n = 205) | | | | | | |
| | | \| (β group 1 - β group 2) \| | | t-value | Sig | p-value |
| St-1 V. St-2 | t1/t2 | 0.159 | | 7.516 | Sig | 0.001 |

**Table 2. Longitudinal Assessment**

Combining the structural model and multi-group analyses, we employed the *"autoregressive effects in latent growth curve modeling, a covariance-based approach to model longitudinal data"* (Roemer 2016, p. 1903). These auto-regressive effects relate to the stability of the constructs from one point in time to the next point in time (Duncan et al. 2013). Here, we created a single PLS model, with constructs at different times (See Table 2). The investigation revealed that the SRMR of the model was 0.041, less than the heuristic of 0.08 (Dijkstra and Henseler 2015) demonstrating sufficient fit. Once the measurement model criteria were assessed and fulfilled, the structural model was evaluated using $R^2$ (same bootstrapping as before). Results of a 'sizeable' effect demonstrated in Table 2 means that the individuals' estimation of the construct remains stable over time (Johnson et al. 2006).

## Conclusions

A unique opportunity in relation to a mandated temporary ban of social media provides the context to this study. Conducted within the premise of social media use as an addiction, the background of the social media ban allowed the researchers to understand how such an addiction transpose itself and its effects (i.e. psychological distress). In addition, the study observed how one transfers from being a 'regular user' to 'non-user' and then revert to a 'regular user.' The study employed a longitudinal design that included two surveys; the first survey was executed during the social media lockdown and the second was completed several days after the ban was lifted. While we acknowledge the wealth of possibilities for research in the context of use and non-use, this study focused only on the psychological distress of such a transition.

The preliminary results reported herein confirm that the impact of social media addiction on psychological distress increases when the use of social media is banned. Literature on addictive behaviors and psychology of addiction provides a possible explanation (Savci and Aysan 2017). Such research suggests that, when the addicted substance or condition (in this case social media use) is discontinued, the individual demonstrates psychological distress as a withdrawal symptom (Hawi and Samaha 2017). Not only did the ban of social media use (i.e. non-use) led to psychological distress, our longitudinal analysis demonstrated that the withdrawal conditions arose when non-use continued to affect the levels of psychological distress, even when the ban was lifted (i.e. use). A series of multi-group analyses led to the discovery of several insights on the relationship between social media addiction and psychological distress. Specifically, it demonstrated that the number of social media apps regularly used, number of friends in social media platforms, frequency of social media use and gender (i.e. female) all having significantly stronger influence on the theorized relationship between social media addiction and psychological distress.

This study makes several research contributions. First, this study provided insights from rare and unique scenario of social media non-use, providing insights of a new social media non-user type that is labeled as 'expelled.' Second, the social media addiction was tested in the new context, providing further generalizability. Third, to socio-psychological studies, this study provides evidence of the existence and its transition of social media addiction in a two-phased longitudinal study. Further, to the broader IS research, this study contributes through a methodological and empirical discussion on social media non-use. For the practice, first, this study shows negative consequences of governments meddling with the platforms of information exchange. Second, from a socio-psychological standpoint, policy makers could create intervention strategies for the identified negative consequences of the over reliance of social media, social media addiction, and potential mental harm of social media. There are several limitations of the study. First, the study did not have an opportunity to gather data in the pre-lockdown phase, which would have provided further evidence of the transition of social media addiction. Second, Study-2 data sample is relatively small for further in-depth analysis. For example, in addition to the multi-group analyses reported here, the study could have included social media app specific analysis (e.g., Facebook vs. SnapChat) and observe the effects of such factors like religious orientation and political orientation. Such nuanced analyses would explain why psychological distress continues, even after the social media ban was lifted. There are several extensions of the study that are currently underway. There have been over 200 interviews that were completed documenting the experience that individuals faced in two study periods. Using latent Dirichlet allocation (LDA) (Blei et al. 2003), the study expects to extend the findings to such aspects like the perceived increased value of social media during and post lockdown and the process of returning to use social media.